\documentclass[apjl,appendixfloats]{emulateapj}
\usepackage{amsmath}
\usepackage{graphicx}
\usepackage{color}
\usepackage{epsfig}
\newcommand{\K}{{\rm K}}

\newcommand{\msun}{{\rm M}_\odot}
\newcommand{\rsun}{R_\odot}
\newcommand{\zsun}{Z_\odot}
\newcommand{\cc}{{\rm cm}^{-3}}

\newcommand{\msunyr}{M_\odot~{\rm yr}^{-1}}

\slugcomment{}

\shorttitle{TDEs by DCBHs}
\shortauthors{K. Kashiyama \& K. Inayoshi}

\begin{document}

\title{Stellar Tidal Disruption Events by Direct Collapse Black Holes}

\author{Kazumi Kashiyama\altaffilmark{1}}

\author{Kohei Inayoshi\altaffilmark{2}}

\affil{$^1$Einstein Fellow--Theoretical Astrophysics Center, Department of Astronomy, University of California, Berkeley, CA 94720, USA}
\affil{$^2$Simons Society of Fellows--Department of Astronomy, Columbia University, 550 West 120th Street, New York, NY 10027, USA}

\email{$^1$ kashiyama@berkeley.edu}
\email{$^2$ inayoshi@astro.columbia.edu}

\begin{abstract}
We analyze the early growth stage of direct-collapse black holes (DCBHs) with $\sim 10^{5} \ \rm M_\odot$, which are formed by collapse of supermassive stars in atomic-cooling halos at $z \gtrsim 10$. 
A nuclear accretion disk around a newborn DCBH is gravitationally unstable and fragments into clumps with a few $10 \ \rm M_\odot$ at $\sim 0.01-0.1 \ \rm pc$ from the center. 
Such clumps evolve into massive population III stars with a few $10-100 \ \rm M_\odot$ via successive gas accretion and a nuclear star cluster is formed.
Radiative and mechanical feedback from an inner slim disk and the star cluster will significantly reduce the gas accretion rate onto the DCBH within $\sim 10^6 \ \rm yr$.
Some of the nuclear stars can be scattered onto the loss cone orbits also within $\lesssim 10^6 \ \rm yr$ and tidally disrupted by the central DCBH.  
The jet luminosity powered by such tidal disruption events can be $L_{\rm j} \gtrsim 10^{50} \ \rm erg \ s^{-1}$. 
The prompt emission will be observed in X-ray bands with a peak duration of $\delta t_{\rm obs} \sim 10^{5-6} \ (1+z) \ \rm s$ followed by a tail $\propto t_{\rm obs}^{-5/3}$, 
which can be detectable by {\it Swift} BAT and eROSITA even from $z \sim 20$. 
Follow-up observations of the radio afterglows with, e.g., eVLA and the host halos with JWST could probe the earliest AGN feedback from DCBHs. 
\end{abstract}


\section{Introduction}
In the last decades, high-z quasar surveys have discovered supermassive black holes~(SMBHs) 
with $M_\bullet \gtrsim 10^{9} \ \rm M_\odot$ at $z > 6$ \citep[e.g.][]{Mortlock_2011,Wu_2015},
which pose questions about the earliest coevolution of SMBHs and their host galaxies.
The observations indicate that at least a minor fraction of high-z BHs experience an extremely efficient mass gain.  
Several scenarios have been proposed~\citep[e.g.,][and references therein]{2012Sci...337..544V,2013ASSL..396..293H}. 
In order to discriminate them, it is important to clarify observational signatures of rapidly growing high-z BHs in the light of the capability of ongoing and upcoming facilities.

Here, we focus on high-z SMBH formation through collapse of supermassive stars~(SMSs) with $\gtrsim 10^{5} \ \rm M_\odot$.  
SMSs can be formed from pristine gas in the so called atomic-cooling halo 
where radiative cooing by atomic hydrogen (H) is relevant instead by molecular-hydrogen (H$_2$)~\citep{BL_2003,Begelman_2006,LN_2006}. 
Formation of H$_2$ can be suppressed by Lyman-Werner and H$^-$ photodissociation photons 
from nearby star-forming galaxies \citep{Omukai_2001,SBH_2010,Visbal_2014,Sugimura+14,Agarwal+15},
and/or collisional dissociation triggered by galactic shocks~\citep{Inayoshi_Omukai_2012,Inayoshi_et_al_2015}.
Once $\gtrsim 10^{5} \ \rm M_\odot$ of such a gas is assembled, it becomes Jeans unstable and collapses to form 
a supermassive protostar at the center~\citep{IOT14,Becerra_2015,Latif_2015}. 
After $\sim \rm Myr$ of gas accretion, the SMS becomes as massive as $\sim 10^5 \ M_\odot$~\citep{Hosokawa_2012,Hosokawa_2013,IHO_13} and  
can directly collapse into a SMBH due to the general-relativistic instability \citep{Iben_1963,Chandrasekhar_1964,Baumgarte_Shapiro_1999,Shibata_Shapiro_2002,Reisswig_et_al_2013}. 
Such direct-collapse black holes (DCBHs) are an attractive candidate of seeds of observed high-$z$ SMBHs.

\begin{figure*}
\begin{centering}
\includegraphics[width=160mm]{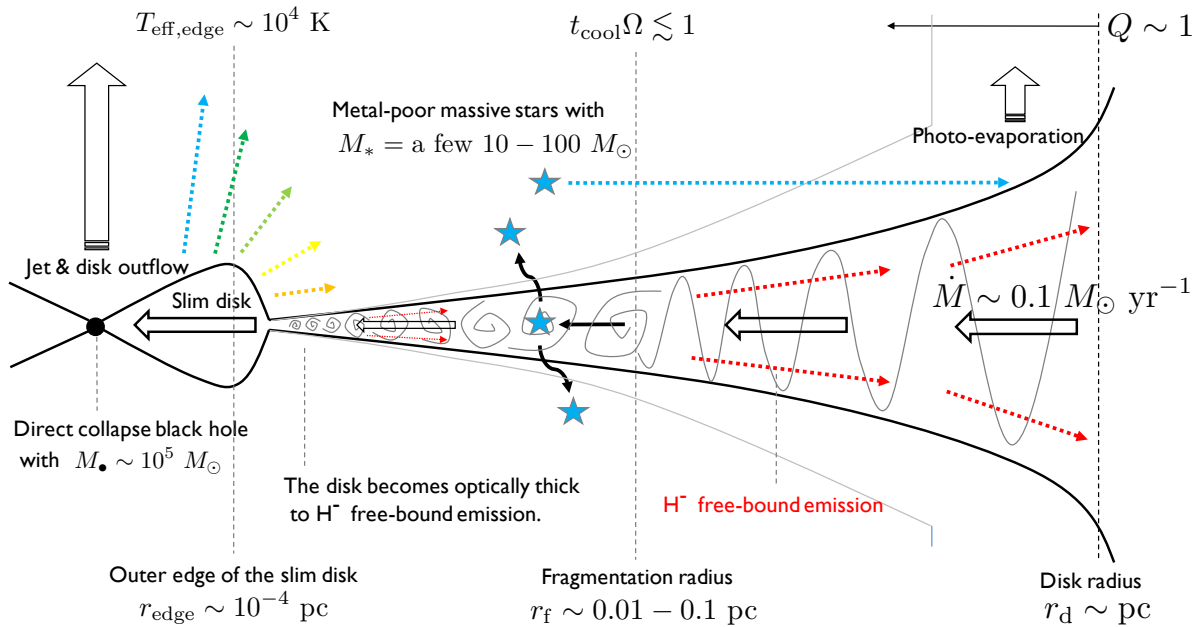}
\caption{Schematic picture of a nuclear accretion disk of a direct collapse black hole within a few Myr after its formation.  
}
\label{mdot}
\end{centering}
\vspace{2\baselineskip}
\end{figure*}

In the very early growth stage of a DCBH, gas accretion onto the nascent DCBH will result in feedback on the surrounding medium, 
which might be probed by James Webb Space Telescope~\citep[JWST;][]{Gardner+06} or future 30-40 meter class telescopes, as argued in e.g.,~\cite{Johnson_et_al_2011}. 
However, given the field of view ($2.2 \times 4.4 \ \rm arcmin^2$ for JWST NIRCam), 
a blind search is not necessarily an efficient way to identify newborn DCBHs and their host halos because DCBH formation may be rare in the early Universe.  
Also we note that X-ray emission from steady accretion onto DCBHs may be challenging to detect even with Chandra~\citep[e.g.,][]{Hartwig+15}.
An alternative strategy is to first find candidates by some transient signatures with survey facilities e.g., {\it Swift} BAT~\citep{Gehrels_et_al_2004} and then to do followup observations with e.g., JWST. 
For example, \cite{Matsumoto_et_al_2015,Matsumoto_et_al_2015b} argued that ultra-long gamma-ray bursts may accompany DCBH formation.

Here, we propose that tidal disruption events (TDEs) of stars formed in the nuclear accretion disk by a newborn DCBHs could occur.  
As we discuss later, associated X-ray bursts and radio afterglows may be much brighter than TDEs in the Universe today and could be detected 
by all-sky instruments such as BAT and eROSITA (Merloni et al. 2012) and on-going radio facilities ike the Expanded Very Large Array (eVLA) even from $z \sim 20$.

This paper is organized as follows. 
In Sec. \ref{sec:disk}, we analyze a nuclear accretion disk around a newborn DCBH. 
First we study disk fragmentation and massive population III (Pop III) star formation in the outer disk, and then discuss properties of the inner slim disk.
In Sec. \ref{sec:tde}, we show that TDEs of the massive Pop III stars by the central DCBH are feasible in the early growth stage and consider the observational signatures. 
We summarize our paper in Sec. \ref{sec:summary}.

\vspace{1\baselineskip}
\section{Nuclear accretion disk around \\ direct collapse black holes}
\label{sec:disk}
We consider a H$_2$-free gas cloud with $\gtrsim 10^{5} \ M_\odot$ collapsing isothermally by H cooling (Ly$\alpha$, two-photon, H$^-$ free-bound, free-free emission) in an atomic-cooling halo with typically $\sim 10^{7-8} \ M_\odot$. 
First, a proto-SMS with $\sim 1-10 \ M_\odot$ and a radius of $\sim 10^{3} \ R_\odot$ is formed at the core~\citep{IOT14,Becerra_2015,Latif_2015}.  
The gas accretion rate onto the core is very high;  
\begin{equation}\label{eq:M_dot}
\dot M \approx \frac{c_{\rm s}^3}{G} \sim 0.095~\msunyr \left(\frac{T_{\rm vir}}{8000~\K}\right)^{3/2}, 
\end{equation}
where $c_{\rm s} = (k_{\rm B}T/\mu m_{\rm p})^{1/2}$ is the sound velocity. 
As for the mean molecular weight, we set $\mu = 1.22$ assuming the pristine abundance (see Sec. \ref{sec:metal} for the effect of metal and dust). 
Note that the above accretion rate is conservative and can be larger than $\sim 1\ M_\odot \ \rm yr^{-1}$
for massive atomic-cooling halos~\citep{Wise+08,Regan_Haehnelt_09b,Regan_Haehnelt_09a}.
The proto-SMS grows via accretion through a massive self-gravitating disk with a size of $r_{\rm d}\sim 1$ pc \citep[e.g.,][]{Regan_2014}, 
to become a SMS with $\sim 10^{5} \ M_\odot$~\citep[e.g.,][]{Hosokawa_2012,Hosokawa_2013,IHO_13,Schleicher_2013,Inayoshi_Haiman_2014}, 
and then, to collapse into a DCBH~\citep[e.g.,][]{Shibata_Shapiro_2002,Reisswig_et_al_2013}. 
The total gas mass in the atomic cooling halo is sufficient to supply gas onto the nuclear disk at a rate as high as Eq. (\ref{eq:M_dot}) 
even after the DCBH is formed unless radiative feedback has been turned on (see Sec. \ref{sec:irr}). 

\subsection{Outer thin disk}
\label{sec:outer}

The outer part of the nuclear accretion disk will be self-gravitating and the gas angular momentum can be transferred mainly by turbulent viscosity due to gravitational instability.
The turbulence also heats the disk and can keep the disk marginally stable, i.e., the Toomre parameter
\begin{equation}\label{eq:Q}
Q \equiv \frac{c_{\rm s}\Omega}{\pi G \Sigma} \sim 1,
\end{equation}
as long as the cooling time of the disk is sufficiently longer than the dynamical time. 
Here, $\Omega$ is the orbital frequency of the disk and $\Sigma$ is the disk surface mass density, which is given by 
\begin{equation}\label{eq:Sigma}
\Sigma = \frac{\dot M}{3 \pi \nu}. 
\end{equation}
We parameterize the disk viscosity as  
\begin{equation}\label{eq:nu}
\nu = \alpha c_{\rm s} h
\end{equation}
\citep{Shakura_Sunyaev_1973}, where 
\begin{equation}\label{eq:h}
h=c_{\rm s}/\Omega 
\end{equation}
is the disk scale height.  
The equation of the heat balance is given by 
\begin{equation}
\frac{9}{4} \nu \Sigma \Omega^2 = 2 h \Lambda_{\rm H^{-}},
\end{equation}
where the left and right hand side represents the viscous heating and radiative cooling, respectively.  
As for the radiative cooling rate (in units of ${\rm erg \ s^{-1} \ cm^{-3}}$), 
we use an approximate formula~\citep{Inayoshi_Haiman_2014}\footnote{
We set the mean molecular weight of the gas to $\mu = 1.22$.  This makes the disk fragmentation radius a factor smaller than that obtained by \cite{Inayoshi_Haiman_2014} with $\mu = 2.0$.};  
\begin{equation}\label{eq:Lambda}
\Lambda_{\rm H^-} \approx 5.0 \times 10^{-41}~T^{2.2}~n^{5/2} e^{-\frac{1.27 \times 10^5}{2T}},  
\end{equation}
where $T$ is the disk temperature in unit of Kelvin and 
\begin{equation}\label{eq:n}
n = \frac{\Sigma}{\mu m_{\rm p} (2 h)}.
\end{equation}
is the number density of hydrogen nuclei in unit of $\rm cm^{-3}$. 
Eq. (\ref{eq:Lambda}) is valid for dense ($n \gtrsim 10^8 \ \rm cm^{-3}$) and warm ($3000 \ {\rm K} \lesssim T \lesssim 8000 \ \rm K$) pristine gas, which is predominantly neutral and optically thin, 
and the main cooling process is free-bound emission of H$^{-}$ ions~\citep[${\rm H}+{\rm e}^-\rightarrow {\rm H}^-+\gamma$,][]{Omukai_2001}. 
From Eqs. (\ref{eq:Q}-\ref{eq:n}), we can obtain the radial profile of the outer disk, $\Sigma$, $h$, $T$ and $\alpha$ for a fixed ($\dot{M}$, $Q$) 
as a function of $\Omega$ or distance from the center, $r = (G M_\bullet/\Omega^2)^{1/3}$, 
where $M_\bullet$ is the mass of the DCBH.

\subsubsection{Disk fragmentation}
\label{sec:clump}

Once the cooling time and dynamical time becomes comparable, 
the disk will fragment into clumps at around a characteristic radius $r_{\rm f}$~\citep[e.g.,][]{Shlosman_Begelman_1987,Shlosman_Begelman_1989,Goodman_Tan_2004,Levin_2007,
Inayoshi_Haiman_2014,Latif_et_al_2015}. 
The cooling time of the disk is estimated as 
\begin{equation}
t_{\rm cool} \approx \frac{\Sigma c_s^2/h}{\Lambda_{\rm H^-}} = \frac{8}{9} (\alpha \Omega)^{-1}.
\end{equation}
On the other hand, from Eqs. (\ref{eq:Q}-\ref{eq:h}), the effective $\alpha$ parameter is given by
\begin{equation}
\alpha = \frac{G\dot{M}Q}{3c_{\rm s}^3} \propto T^{-3/2}.
\end{equation}
In the outer disk, the gas temperature gradually decreases toward the center, 
the value of $\alpha$ increases inward and finally exceeds a critical value (i.e. $t_{\rm cool}\Omega \la1$), 
which has been numerically calculated to be $\alpha_{\rm f} \lesssim 1$~\citep[e.g.,][]{Gammie_2001,Rice_et_al_2003,Krumholz_2007,Clark_2011,Zhu_2012}. 
For our fiducial parameter set ($\dot{M} = 0.095 \ \rm M_\odot \ yr^{-1}$, $Q =1.0$, $\alpha_{\rm f} =1.0$, $M_\bullet = 1.0\times 10^{5} \ M_\odot$), 
the disk fragmentation occurs at 
\begin{equation}\label{eq:rf}
r_{\rm f} \sim 3.3\times 10^{-2}~{\rm pc},
\end{equation}
where the orbital period is $t_{\rm orb,f}=2\pi /\Omega_{\rm f}\simeq 1.8\times 10^3$ yr.
We find that the disk properties around the fragmentation radius are 
\begin{equation}
\Sigma \sim 270~{\rm g \ cm^{-2}} \left(\frac{r}{r_{\rm f}}\right)^{-3/2},
\end{equation}
\begin{equation}
n \sim 1.4 \times 10^{10}~{\rm cm^{-3}} \left(\frac{r}{r_{\rm f}}\right)^{-3}, 
\end{equation}
\begin{equation}
h/r \sim 0.045 \left(\frac{r}{r_{\rm f}}\right)^{1/2},
\end{equation}
\begin{equation}\label{eq:T}
T \sim 3800~{\rm K} \left(\frac{r}{r_{\rm f}}\right)^{-0.16}. 
\end{equation}
Typical mass of the clumps can be estimated as
\begin{equation}\label{eq:Mc0}
M_{\rm c, 0} \approx (2 \pi h_{\rm f})^2 \Sigma _{\rm f} \sim 28~\msun,
\end{equation}
where $2 \pi h_{\rm f}$ corresponds to the wavelength of the most unstable mode.

After fragmentation, the clumps grow in mass and migrate inward by interacting with the disk. 
The mass accretion rate onto the clump can be estimated as 
\begin{equation}\label{eq:dotMc}
\dot M_{\rm c} = \frac{3}{2} \Sigma_{\rm f} \Omega_{\rm f} (f_{\rm H} R_{\rm H})^2  \sim 1.6 \times 10^{-2}~f_{\rm H}^2~\msunyr ,
\end{equation}
where $R_{\rm H} = r_{\rm f} (M_{\rm c, 0}/3M_\bullet)^{1/3}$ is the Hill radius of the clump and $f_{\rm H} \sim O(1)$.
The initial migration time of the clump, during which the rotation radius typically decreases to $\sim r_{\rm f}/2$, 
can be approximately given by Type I migration time~\citep{Zhu_2012}, 
\begin{equation}\label{eq:tmig}
t_{\rm mig} \approx \frac{1}{4 C q_{\rm f} \mu_{\rm f}} \left(\frac{h_{\rm f}}{r_{\rm f}}\right)^2 \frac{2 \pi}{\Omega_{\rm f}} \sim 1.3 \times 10^4 \ \rm yr, 
\end{equation}
where $C = 3.2 + 1.468 \ \xi$ with $\xi \approx 1.5$ being the power index of the surface density, 
$q_{\rm f} = M_{\rm c,0}/M_\bullet$, and $\mu_{\rm f} = \pi \Sigma_{\rm f} r_{\rm f}^2/M_\bullet$~\citep{Tanaka_et_al_2002}. 
From Eqs. (\ref{eq:dotMc}-\ref{eq:tmig}), the total clump mass could increase to 
$M_{\rm c} \sim 200 \ M_\odot$ during the migration time ($\sim 10~t_{\rm orb,f}$).
This total mass is comparable to the isolation mass~\citep[e.g.,][]{Goodman_Tan_2004};
\begin{equation}
M_{\rm iso} \approx \frac{(2 \pi f_{\rm H} \Sigma r^2)^{3/2}}{9 M_\bullet^{1/2}} \sim 180~f_{\rm H}^{3/2} \left(\frac{r}{r_{\rm f}/2}\right)^{3/4} M_\odot, 
\end{equation}
and a gap can be formed around the clump. 
This also means that a reasonable fraction of the gas at the fragmentation radius can accrete onto at most a few clumps.
Typical clump mass at $\sim r_{\rm f}/2$ will range from $M_{\rm c} = {\rm a \ few} \times (10-100) \ M_\odot$,  
which is broadly consistent with numerical simulations of a self-gravitating disk around a SMS~\citep{Sakurai_2015b}. 

\subsubsection{Star formation\footnote{
\cite{Aykutalp+14} discussed Pop III star formation in halos hosting a DCBH but at greater distances from the center.}}
\label{sec:star_formation}

Just after the fragmentation, the central core of the clump collapses in a runaway fashion via optically-thin H$^{-}$ free-bound emission.   
The core finally becomes optically thick and a quasi-hydrostatic equilibrium state, i.e., a protostar is formed, within a free-fall time of $\sim 500~{\rm yr}~(n_{\rm f}/10^{10}~\cc)^{-1/2}$.
During the migration time, the clumps grow via gas accretion at a rate of $\sim 2\times 10^{-2}~\msunyr$ (Eq. \ref{eq:dotMc}),
which is higher than a critical value, $\dot{M}_{\rm crit} \sim 4\times 10^{-3}~\msunyr$ \citep{Omukai_Palla_2003}. 
In this case, the protostellar structure consists of two parts; 
a bloated convective envelope with a radius of $R_{\rm c} \sim 100~(2000)~ \rm R_\odot$ for $M_{\rm c} \sim 10~(100) \ \rm M_\odot$ and a radiation dominated core which dominates the mass~\citep{Hosokawa_2012,Hosokawa_2013}. 

The core contract within a Kelvin-Helmholtz (KH) time; 
\begin{equation}
t_{\rm KH} \approx 3600  \ {\rm yr} \ \beta_\ast^{-0.053} (1 - \beta_\ast)^{0.89}, 
\end{equation}
and evolves to a Pop III main-sequence star with a mass of 
\begin{equation}\label{eq:Mast}
\frac{M_\ast}{M_\odot} \approx 52 \frac{\sqrt{1-\beta_\ast}}{\beta_\ast^2}, 
\end{equation}
and a radius of 
\begin{equation}\label{eq:Rast}
\frac{R_\ast}{R_\odot} \approx 4.5 \frac{(1-\beta_\ast)^{0.39}}{\beta_\ast^{0.95}}, 
\end{equation}
where $\beta_\ast$ is the dimensionless parameter roughly corresponding to the ratio of the gas pressure to the total pressure~\citep{Goodman_Tan_2004}. 
For example, we can estimate $t_{\rm KH}=1.3~(0.74)\times 10^4$ yr, $M_\ast = 40~(100)~\msun$, and $R_\ast = 3.1~(5.3)~\rsun$. 
After a migration time (Eq. \ref{eq:tmig}), the clump mass is close to the isolation mass of the disk and a gap will be formed.  
Then, gas accretion onto each clump will significantly decrease, which sets the maximum mass of the star to be a few $100 \ M_\odot$.  
We note that, although multiple stars can be formed from a single clump, the mean stellar mass is not much smaller than the total clump mass 
if the mass function is top-heavy, which is likely in the case of star formation from metal-poor gas \citep[e.g.][]{Hirano_2014}.  

The main-sequence time of stars with $\langle M_\ast \rangle \sim 30-100 \ M_\odot$ is 
\begin{equation}\label{eq:t_life}
t_{\rm MS} \approx \frac{\varepsilon_{\rm nuc} M_\ast c^2}{L_\ast} \sim (2-4) \ \rm Myr,
\end{equation}
where $\epsilon_{\rm nuc}\sim 0.01$ is the energy conversion rate in nuclear burning and $L_\ast$ is close to the Eddington luminosity~\citep{Schaerer_2002}. 
Massive Pop III stars with a sufficient rotation would evolve into red supergiants after the main-sequence phase. 
This is primarily due to a rotational and convective mixing of CNO elements into the outer envelope~\citep[e.g.,][]{Hirschi+07,Ekstr+08}. 
Finally, such massive Pop III stars collapse and supernova explosions occur, leading to metal enrichment in the nuclear region of the host halo, 
although the outcome will be sensitive to the stellar mass, rotation, and magnetic field~\citep[e.g.,][]{Joggerst+10,Joggerst+11,Yoon+12}. 
Since we are interested in a few Myr after the DCBH formation, which is comparable to the main-sequence lifetime (Eq. 23), hereafter we only consider the main-sequence phase.

\subsection{Inner slim disk}
\label{sec:inner}
In the inner region, the disk finally becomes optically thick to H$^{-}$ free-bound absorption when $\alpha_{\rm H^{-}} r \ga 1$,
which is $r \la 2.6 \times 10^{-4} \ \rm pc$ for our fiducial case. 
Here, $\alpha_{\rm H^{-}} = 4 \pi \Lambda_{\rm H^{-}}/B(T)$ is the absorption coefficient and $B(T) = \sigma_{\rm SB} T^4/\pi$. 
Note that we take $r$ instead of $h$ as the path length of the photon. 
The photons can escape mainly in the radial direction since optically thick ionized layers are formed above the thin disk 
due to the irradiation as we show in the next subsection. 

Once the disk becomes optically thick to H$^{-}$ free-bound absorption, both of the temperature and ionization fraction of the disk start to increase due to viscous heating. 
The accretion rate is super-Eddington with respect to Thomson scattering in fully ionized gas, $\dot{M} \gtrsim 500~L_{\rm Edd}/c^2$. 
Thus, once the disk is significantly ionized, the inner disk becomes the so-called slim disk~\citep{Abramowicz_et_al_1988,Beloborodov_1998}. 
The scale height of the slim disk takes its maximum at a radius where the radiation pressure dominates the gas pressure. 
In our case, this occurs at 
\begin{equation}
r_{\rm edge} \approx \left[ \frac{3(1-\eta_\ast)GM_\bullet \dot{M}}{8 \pi \sigma_{\rm SB} T_{\rm edge}^4}\right]^{1/3} \sim 10^{-4} \ \rm pc, 
\end{equation}
with $T_{\rm eff, edge} \sim 10^{4} \ \rm K$ being the effective temperature at the outer edge of the inner disk.
In $r < r_{\rm edge}$, the temperature scales as $T_{\rm eff} \approx T_{\rm eff, edge} (r/r_{\rm edge})^{-1/2}$ and the aspect ratio of the disk becomes $h/r \gtrsim 0.3$. 
We note that ionization instability can occur in this transition region between the outer thin and inner slim disk, which may result in an episodic accretion~\citep[e.g.,][]{Lin_Shields_1986}. 

Given that the accretion rate is well above the Eddington rate, a radiation-driven wind will be launched from the inner disk. 
Also, if large-scale magnetic fields exist in the innermost region, a relativistic jet can be launched by the Blandford-Znajek mechanism~\citep{Blandford_Znajek_1977}.
The total bolometric luminosity can be as large as $L_{\rm AGN} \approx \eta_{\rm AGN} (1 - \eta_\ast) \dot M c^2$, or
\begin{equation}
L_{\rm AGN} \sim 7.4 \times 10^{44} \ \eta_{\rm AGN} (1 - \eta_\ast) \ \rm erg \ s^{-1}. 
\end{equation}
The efficiency $\eta_{\rm AGN}$ is estimated to be $\sim 10 \%$ for the disk wind~\citep[e.g.,][]{Ohsuga_et_al_2005,Jiang_et_al_2014,Sadowski_et_al_2014} 
and can be even larger for the jet~\citep{Tchekhovskoy_et_al_2011}. 
Such disk wind and jet can give significant radiative and mechanical feedback on the parent halo, especially in the polar region.
Although direct emission from the wind and jet are not detectable from $z \gtrsim 10$, the feedback effects can be indirectly probed by observing reprocessed emission, 
in particular H$\alpha$ and HeII $\lambda$1640 with JWST~\citep[e.g.,][]{Johnson_et_al_2011}. 

\subsection{The effects of irradiation}
\label{sec:irr}

Here, we consider the effect of irradiation of EUV photons from the inner region on the outer neutral disk.
Inside a critical radius, $r < r_{\rm pe} \approx G M_\bullet/c_{\rm s, HII}^2 \sim 0.67 \ {\rm pc} \ (M_\bullet/10^5 \ M_\odot)$, 
optically-thick ionized layers with a scale height of $h_{\rm HII} \approx c_{\rm s, HII}/\Omega$ is formed above the outer disk. 
Here, $c_{\rm s, HII} \sim 20 \ \rm km \ s^{-1}$ is the sound speed in the ionized region. 
Outside the critical radius, $r_{\rm pe} < r < r_{\rm d}$, the disk is photo-evaporated by the irradiation.
The evaporation rate is estimated as 
$\dot{M}_{\rm pe} \sim 6.6 \times 10^{-2} \ {\rm M_\odot \ yr^{-1}} \ (\Phi/10^{50} \ {\rm s^{-1}})^{1/2} (r_{\rm d}/{\rm pc})^{1/2}$, where $\Phi$ is the ionizing photon number,
and gas supply from large scales onto the outer disk can be suppressed when 
$\dot{M}_{\rm pe}/\dot{M} \gtrsim 0.2$~\citep{Tanaka_et_al_2013}.
The critical ionization photon number flux at the disk radius is 
$\Phi_{\rm crit} \sim 9.5 \times 10^{50} \ {\rm s^{-1}} \  (r_{\rm d}/{\rm pc})^{-1}(\dot{M}/0.1 \ M_\odot \ {\rm yr^{-1}})^2$.

Before the direct collapse occurs, the main irradiation source is the central SMS. 
The accretion disk merges to the SMS at the surface, $R_{\rm sms} \sim 3 \times 10^{-4} \ \rm pc$. 
The surface temperature of the SMS can be as high as $T_{\rm sms} \sim 10^4 \ \rm K$ just before the direct collapse. 
The ionization photon flux can be $\Phi_{\rm sms} \sim 10^{50} \ \rm s^{-1}$~\citep{Hosokawa_2013}, 
which is still below the critical value. 

Once the DCBH is formed, the accretion disk extends inward to the BH horizon scale. 
High-energy photons emitted from the inner slim disk are shielded by the slim disk itself. 
Instead, the outer disk is mainly irradiated by the outer edge of the slim disk, where $T_{\rm edge} \sim 10^{4} \ \rm K$ and $r_{\rm edge} \sim 10^{-4} \ \rm pc$ (see Fig. \ref{mdot} and \S\ref{sec:inner}).
The corresponding ionization photon number flux is $\Phi_{\rm edge} \sim 10^{49} \ \rm s^{-1}$, which is also below the critical value. 

Finally, massive Pop III stars formed by disk fragmentation can be an important irradiation source.  
The ionization flux from each star is $\Phi_\ast \sim 2 \times 10^{49} \ {\rm s}^{-1} \ (M_\ast/40 \ M_\odot)^{3/2}$. 
About $\sim$Myr after the DCBH formation, the total stellar mass becomes $\approx \eta_\ast M_\bullet$, where $\eta_\ast = 0.1$ is the star formation efficiency, 
and the number of stars can be $N_\ast \approx \eta_\ast M_\bullet/\langle M_\ast \rangle \sim 250$.
Then, the total ionization flux can be $\Phi_{\rm cluster} \approx N_\ast \Phi_\ast \sim 5 \times 10^{51} \ \rm s^{-1}$, which is larger than the critical value. 
Thus, we can concluded that the radiative feedback from the nuclear star cluster can significantly suppress 
the disk accretion rate after a mass doubling time of the DCBH. 

We note that massive stars also ionize the gas within the disk thickness,
but the size of the ionized regions will be much smaller than the disk radius~\citep{Inayoshi_Haiman_2014}. 
The outer thin disk is still predominantly neutral except near the disk surface even after the star formation. 
Thus our assumption in Sec. \ref{sec:outer} is justified.

\subsection{The effect of metal and dust}\label{sec:metal}

In Sec. \ref{sec:outer}, we assume that the disk consists of pristine gas,
and consider only H$^{-}$ emission as the radiative cooling process.  
Here, we discuss whether this assumption would be valid for a low-metallicity environment as well.

Even at $z \gtrsim 10$, the metallicity in an atomic-cooling halo could be as high as $Z \sim 10^{-3}-10^{-4} \ Z_\odot$ 
due to past supernova explosions \cite[e.g.][]{Bromm+03,Whalen+08,Greif_2010,Wise_2012,Ritter_2012},  
although the uncertainties are fairly large depending on the evolution history of each halo 
and on stellar progenitors which occur the explosions. 
We can neglect metal line cooling in the atomic-cooling halo as long as $Z \lesssim 10^{-3} \ Z_\odot$. 
On the other hand, thermal emission from dust grains could be an important coolant even if the metallicity is as low as $\sim 10^{-4}~\zsun$ \citep{Omukai_2008}. 

In a low density region in the atomic cooling halo, the dust and gas are essentially decoupled. 
Due to the efficient emission, the dust temperature is kept much lower than surrounding gases, $T_{\rm d} \sim 100 \  \rm K$. 
When the gas density increases toward the center of the nuclear disk, 
the dust and gas start to exchange heat by collisions within the dynamical time, 
which occurs when
\begin{equation}
n \gtrsim 3 \times 10^{9} \ {\rm cm^{-3}} \left(\frac{T}{4000 \ \rm K}\right) \left(\frac{f_{\rm dep}}{0.1}\right)^{-2} \left(\frac{Z}{10^{-4} \ Z_\odot}\right)^{-2}, 
\end{equation}
\citep{Schneider_2006,Schneider_2012}, where $f_{\rm dep}$ is the dust depletion factor.  
From Eqs. (\ref{eq:n}) and (\ref{eq:T}), the above condition is realized at $r \lesssim r_{\rm f}$ for $Z \sim 10^{-4} Z_\odot$.

On the other hand, dust grains are sublimated once the dust temperature exceed a critical value of $\sim 10^3~\K$.
Considering the inner slim disk as the main irradiation source, the dust temperature is estimated as $T_{\rm d} \approx T_{\rm edge} (r/r_{\rm edge})^{-1/2}$.
Therefore, the effect of dust cooling could be neglected within $\sim 0.01 \ \rm pc$, 
which is close to the typical star formation radius, $r \sim r_{\rm f}/2$.  

From the above arguments, we conclude that the dust cooling can be neglected if the gas metallicity 
in the atomic cooling halo is as low as $Z \lesssim 10^{-4} \ Z_\odot$. 
Otherwise, the dust cooling becomes important at $r > r_{\rm f}$ and the nuclear accretion disk is more likely to 
fragment into clumps at larger radii than we showed in Sec. \ref{sec:outer}. 
The initial mass function of the stars formed by disk fragmentation would be less top-heavy compared with the metal-zero case~\citep{Omukai_2008}. 
Also, the size of the star cluster would be larger, resulting in a longer relaxation time.  
Thus, these effects could reduce the total event rate of energetic TDEs by DCBHs (see Sec. \ref{sec:tde}).

\vspace{1\baselineskip}
\section{Stellar tidal disruptions by \\ direct collapse black holes}
\label{sec:tde}

\subsection{Sending stars to the tidal radius}
The orbits of the clumps and stars after the initial migration time is highly nonlinear and stochastic~\citep[e.g.][]{Zhu_2012}.
Three-dimension hydrodynamical simulations are required to predict the evolution. 
If the gas-to-star conversion efficiency is not extremely high, interactions between the stars and residual gas disk will be still effective. 
The stars may migrate further inward with a timescale of Eq. (\ref{eq:tmig}), or a viscous timescale once a gap is formed around the star~\citep[e.g.,][]{Lin_Papalouzou_1986,Ward_1997}; 
\begin{equation}\label{eq:t_vis}
t_{\rm vis} \approx \frac{1}{3 \pi \alpha} \left(\frac{r}{h}\right)^{2} \frac{2\pi}{\Omega} \sim 9.5 \times 10^4 \ \rm yr. 
\end{equation}
Such migration may proceed down at least to the radius where the disk mass becomes comparable to the clump mass, i.e., $\sim 10^{-4} \ \rm pc$ for our fiducial case. 
Before the direct collapse, most of the clumps may merge with the central SMS with a stellar radius of $\sim 10^{-3} \ \rm pc$, as observed in numerical simulations~\citep{Sakurai_2015b}. 
On the other hand, after the direct collapse, a dominant fraction of the stars will not directly migrate to the tidal radius of the DCBH, 
$r_{\rm t} = R_\ast (M_\bullet /M_\ast)^{1/3} \sim 4.9 \times 10^{12} \ {\rm cm} \ (M_\bullet/10^5 \rm M_\odot)^{1/3} (M_\ast/40 \rm M_\odot)^{-1/3} (R_\ast/5 R_\odot)$.
Instead, interactions between the clumps and stars finally become more relevant, resulting in forming a nuclear star cluster.  

Motivated by the argument in Sec. \ref{sec:star_formation}, we assume the effective stellar mass $\langle M_\ast \rangle = 40 \ M_\odot$ and size of the cluster $\lesssim r_{\rm f}/2 \sim 0.01 \ \rm pc$, respectively.  
Initially, most of the stars are alined with the disk. 
The relaxation time of the eccentricity can be estimated as~\citep{Stewart_Ida_2000,Kocsis_Tremaine_2011}
\begin{equation}\label{eq:t_relax_disk}
t_{\rm relax, disk} \approx 0.22 \frac{\langle e^2 \rangle^2 M_\bullet^2}{\Omega \langle M_\ast \rangle \Sigma_\ast r^2 \ln \Lambda} \sim 3.7 \times 10^4 {\ \rm yr} \ \left(\frac{N_\ast}{10} \right)^{-1}. 
\end{equation}
Here,  $\langle e^2 \rangle^{1/2} = 0.3$ is the mean eccentricity, $\Sigma_\ast = N_\ast \langle M_\ast \rangle/\pi r^2$ is the surface density of the stars, 
and $\Lambda = \langle e^2 \rangle^{3/2} M_\bullet/\langle M_\ast \rangle$.
The inclination also relaxes with a timescale of $\approx 2~t_{\rm relax, disk}$. 
A dozen of disk stars are formed within $\approx 10 \times \langle M_\ast \rangle/ (\eta_\ast \dot{M}) \sim 4 \times 10^4 \ \rm yr$. 
Hence the disk stars will evolve into a quasi-spherical stellar cluster within $\lesssim 10^5 \ \rm yr$. 
The relaxation time of the spherical cluster can be estimated as 
\begin{equation}\label{eq:t_relax_cluster}
t_{\rm relax, cluster} \approx 0.34 \frac{\sigma^3}{G^2 \rho_\ast \langle M_\ast \rangle \ln\Lambda} \sim 7.7 \times 10^4 \ \rm yr,
\end{equation}
As for Eq. (\ref{eq:t_relax_cluster}), we consider non-resonant two-body interaction~\citep{Binney_Tremaine_2008}, 
substituting $\sigma = (GM_\bullet/2r)^{1/2}$, $\rho_\ast = \eta_\ast M_\bullet/(4\pi r^3/3)$, and $\ln\Lambda = \ln(M_\bullet/\langle M_\ast \rangle) \sim 7.8$. 
Some stars in the cluster are scattered onto the loss cone orbits within 
\begin{equation}\label{eq:t_tde}
t_{\rm tde} \approx t_{\rm relax, cluster} \times \ln(2/\theta_{\rm lc}) \sim 3.8 \times 10^5 \ \rm yr, 
\end{equation}
where $\theta_{\rm lc}^2 = (r_{\rm t}/r)\times (GM_\bullet/\sigma^2)$ is the angular size of the loss cone~\citep[e.g.,][]{Syer_Ulmer_1999}. 
From Eqs. (\ref{eq:t_life}) and (\ref{eq:t_tde}), TDEs can occur within the lifetime of a massive Pop III star. 
Namely, we can expect $\approx t_{\rm MS}/t_{\rm tde} \sim 10$ of TDEs by each DCBH. 
We note that Eq. (\ref{eq:t_relax_cluster}) is derived assuming a continuous distribution function of stars. 
In our case, the number of stars $N_\ast \lesssim 1000$ may not be large enough, resulting in a relatively large scatter in the number of TDE per DCBH.

\subsection{Observational signatures}
Pop III stars disrupted by newborn DCBHs are massive and compact (see Eqs. \ref{eq:Mast}-\ref{eq:Rast}).  
Consequently, the accretion luminosity of such TDEs can be significantly higher than the low-z events, 
and some associated emissions could be detectable even from high redshifts. 
The fallback accretion rate is estimated as 
\begin{equation}\label{eq:Mdot_fb}
\dot M_{\rm fb} \approx \frac{M_\ast}{3t_{\rm fb}} \left(\frac{t}{t_{\rm fb}}\right)^{-5/3}, 
\end{equation}
where 
\begin{eqnarray}
t_{\rm fb} &\sim& 1.4 \times 10^5 \ {\rm s} \ k^{-3/2} \beta \notag \\ 
&& \times \left(\frac{M_\bullet}{10^5 M_\odot}\right)^{1/2}  \left(\frac{M_\ast}{40 M_\odot}\right)^{-1}  \left(\frac{R_\ast}{3 R_\odot}\right)^{3/2},
\end{eqnarray}
is the fallback time, where $k$ is a factor of $O(1)$ related to structure of disrupted star, $\beta$ is the ratio between the tidal radius and pericenter distance~\citep{Stone_et_al_2013}. 
The peak accretion luminosity is $\dot M_{\rm fb}/\dot M \sim 10^4$ times larger than the mean accretion rate.

If large-scale magnetic fields are amplified during the disruption, a relativistic jet can be launched by the Blandford-Znajek mechanism.
The peak luminosity of the jet is estimated as $L_{\rm j,TDE} \approx \eta_{\rm j} M_\ast c^2/3t_{\rm fb}$, or 
\begin{eqnarray}\label{eq:L_j}
L_{\rm j, TDE} &\sim& 1.7 \times 10^{50} \ {\rm erg \ s^{-1}} \ \eta_{\rm j} k^{3/2} \beta^{-1}  \notag \\
&& \times  \left(\frac{M_\bullet}{10^5 M_\odot}\right)^{-1/2}  \left(\frac{M_\ast}{40 M_\odot}\right)^{2}  \left(\frac{R_\ast}{3 R_\odot}\right)^{-3/2}.
\end{eqnarray}
With the super-Eddington accretion rate associated with TDEs, the innermost disk can be a magnetically arrested disk with an advection-dominated accretion flow~\citep{Tchekhovskoy_et_al_2011}. 
In this case, the jet efficiency can be described as
\begin{equation}
\eta_{\rm j} \sim 1.3 \left(\frac{h/r}{0.3}\right) \left(\frac{a}{M_\bullet}\right)^{2},
\end{equation}
\citep{Tchekhovskoy_2015}.
The Kerr parameter of DCBH is expected to be as large as $a/M_\bullet \sim 0.9$~\citep{Reisswig_et_al_2013}.
Thus, $\eta_{\rm j} \sim 1$ is a viable assumption. 

\subsubsection{Prompt X rays}
Although the prompt emission mechanism of TDE jets is still highly uncertain, 
the peak energy in the engine rest frame will be in hard-X-ray bands~\citep{Bloom_et_al_2011,Levan_et_al_2011,Burrows_et_al_2011,Cenko_et_al_2012}, 
and observed in soft X-ray bands $\lesssim 10 \ \rm keV$ due to redshift.
For an emission efficiency of $10 \%$ and a jet beaming factor of $f_{\rm b} = 0.01$, the peak isotropic luminosity is $L_{\gamma, \rm iso} \sim 8 \times 10^{50} \ \rm erg \ s^{-1}$,
which corresponds to an observed flux of $\sim 6.3 \times 10^{-10} \ \rm erg \ s^{-1} \ cm^{-2}$ from $z = 10$ and $\sim 1.3 \times 10^{-10} \ \rm erg \ s^{-1} \ cm^{-2}$ from $z=20$.  
The observed peak duration is $\delta t_{\rm obs} \sim 10^{(5-6)} \ (1+z) \ \rm s$ for a disrupted star with $M_\ast = 40 \ M_\odot$. 
Such emissions are good targets of soft-X-ray survey telescopes like eROSITA~\citep{Merloni_et_al_2012} and HiZ-GUNDAM~\citep{Yonetoku_et_al_2014}. 
For example, eROSITA has a limiting flux of $\sim 10^{-12} \ \rm erg \ s^{-1} \ cm^{-2}$, though the value depends on the emission spectrum~\citep[e.g.,][]{Khabibullin_et_al_2014}, 
and can detect the above emission even from $z \sim 20$. 
The early tail emission $\propto t_{\rm obs}^{-5/3}$ is also detectable and can be used to distinguish this type of TDEs from other high-z transients~\citep[e.g.,][]{Kashiyama_et_al_2013,Matsumoto_et_al_2015}. 
If the intrinsic spectrum is hard enough, the signal can be also detectable up to $z \sim 20$ by {\it Swift} BAT, but only with an integration time of $\gtrsim 10^4 \ \rm s$~\citep{Baumgartner_et_al_2013}.
A BAT archival data search of dim (in terms of the observed flux) and ultra-long transients will be interesting.  

\subsubsection{Radio afterglow}
A promising counterpart may be the afterglow emission~\citep[e.g.,][]{Ioka_Meszaros_2005,Toma_et_al_2011}. 
The kinetic energy of the jet is as large as $E_{\rm j} \approx L_{\rm j, TDE} t_{\rm fb} \sim 2.4 \times 10^{55} \ {\rm erg} \ (M_\ast/40 \ M_\odot)$.
The jet kinetic energy is comparable or larger than the binding energy of the parent halo.  
Thus, TDEs in collapsing atomic-hydrogen-cooling halos will also give significant mechanical feedback. 
The synchrotron emission from the decelerating jet, especially in radio bands, could be detectable even from $z \sim 20$ by e.g., eVLA as we show below. 
In this subsection, we use the notation $Q = 10^x Q_x$ in CGS unit unless we note.

The typical time scale of the afterglow emission corresponds to the deceleration time of the TDE jet, $t_{\rm dec} \approx (1+z)r_{\rm dec}/4 \Gamma^2 c$, or
\begin{equation}
t_{\rm dec} \sim 1.0 \times 10^{7} \ {\rm s} \ \left(\frac{1+z}{11}\right) f_{\rm b, -2}^{-1} E_{\rm j, 55} \Gamma_{1.3}^{-4} \dot M_{\rm w, -2}^{-1} v_{\rm w, 10} .
\end{equation}
in the observer frame. 
Here, $\Gamma$ is the Lorentz factor of the jet and $r_{\rm dec} \approx E_{\rm j} v_{\rm w}/f_{\rm b}\dot M_{\rm w} \Gamma^2 c^2$, or 
\begin{equation}
r_{\rm dec} \sim 4.4 \times 10^{19} \ {\rm cm} \ f_{\rm b, -2}^{-1} E_{\rm j, 55} \Gamma_{1.3}^{-2} \dot M_{\rm w, -2}^{-1} v_{\rm w, 10},  
\end{equation}
is the deceleration radius. 
We assume a wind-like density profile; $\rho_{\rm w} (r) = \dot M_{\rm w}/4 \pi v_{\rm w} r^2$, 
and assume that $\dot M_{\rm w} = 10^{-2} \ M_\odot \ \rm yr^{-1}$ and $v_{\rm w} = 10^{10} \ \rm cm \ s^{-1}$ as fiducial, 
which corresponds to that a 10 \% of the accreted matter has been ejected from the slim disk with an escape velocity at the inner most region. 
Note that the first TDE typically occurs $t_{\rm tde} \gtrsim 10^{5} \ \rm yr$ after the DCBH formation when the wind region extends far beyond the deceleration radius, $r_{\rm dec} \ll v_{\rm w} t_{\rm tde}$.

At the decelerating shock, magnetic field amplification and electron acceleration can occur. 
The comoving magnetic field strength at the forward shock is estimated as $B \approx [32 \pi \epsilon_{B} \rho_{\rm w}(r_{\rm dec}) c^2 \Gamma^2]^{1/2}$, or
\begin{equation}
B \sim 0.031 \ {\rm G} \ \epsilon_{B, -2}^{1/2}  f_{\rm b, -2} E_{\rm j, 55}^{-1} \Gamma_{1.3}^3 \dot M_{\rm w, -2}^{3/2} v_{\rm w, 10}^{-3/2},
\end{equation}
where $\epsilon_B$ is the magnetic field amplification efficiency. 
We note that the energy density of the cosmic microwave background (CMB) in the comoving frame is much smaller than that of the amplified magnetic field, thus the effect of inverse Compton cooling can be neglected, 
unlike extended lobes produced by steady jets from high-z SMBHs whose radio emission may be muted by the CMB~\citep{Celotti_Fabian_04,Ghisellini+14,Fabian+14}.
The minimum Lorentz factor of the non-thermal electrons is given by 
\begin{equation}
\gamma_{\rm e, m} \approx  \epsilon_{\rm e} \frac{p-2}{p-1} \frac{m_{\rm p}}{m_{\rm e}} \Gamma \sim 1200 \ \epsilon_{\rm e, -1}  \Gamma_{1.3}.
\end{equation}
where $\epsilon_{\rm e}$ is the acceleration efficiency.  
Hereafter we set the power low index of the non-thermal electrons as $p = 2.5$. 
The characteristic synchrotron frequency in the observer frame is $\nu_{\rm m} = \Gamma \gamma_{\rm e, m}^2 q B/[2 \pi m_{\rm e} c (1+z)]$, or
\begin{eqnarray}
\nu_{\rm m} \sim 230 \ {\rm GHz} \ && \left(\frac{1+z}{11}\right)^{-1} \epsilon_{\rm e, -1}^{2} \epsilon_{B, -2}^{1/2}  \notag \\
&& \times f_{\rm b, -2} E_{\rm j, 55}^{-1} \Gamma_{1.3}^6 \dot M_{\rm w, -2}^{3/2} v_{\rm w, 10}^{-3/2},
\end{eqnarray}
and the corresponding flux can be estimated as 
\begin{eqnarray}
F_{\nu, \rm max} &\approx& \frac{\sigma_{\rm T} m_{\rm e} c^2 \dot M_{\rm w} B \Gamma r_{\rm dec} (1+z)}{12 \pi q m_{\rm p} v_{\rm w} D_{\rm L}^2} \notag \\
&\sim& 360 \ {\rm mJy} \ \left(\frac{1+z}{11}\right) D_{\rm L, 29.5}\epsilon_{B, -2}^{1/2} \Gamma_{1.3}^2 \dot M_{\rm w, -2}^{3/2} v_{\rm w, 10}^{-3/2}, \notag \\
\end{eqnarray}
where $D_{\rm L}$ is the luminosity distance to the source. 
For $\nu \lesssim \nu_{\rm m}$, the flux proportional to $\nu^{1/3}$~\citep[e.g.,][]{Granot_Sari_2002}. 
Note that in our fiducial case, the shocked plasma is in the slow cooling regime and synchrotron self absorption is not relevant for $\nu \gtrsim \rm GHz$ at $t \sim t_{\rm dec}$. 
For example, at $\nu \sim 1-10 \ \rm GHz$, the anticipated peak flux $\sim 10^6 \ (1+z) \ {\rm s}$ after the TDE can be $\gtrsim 10$ mJy even from $z \sim 20$, 
which can be detectable by eVLA and SKA.

\subsubsection{Disk emission}
Finally, the quasi-thermal disk emission can be also enhanced by the TDEs. 
In our case, the emission temperature is in EUV or soft-X-ray bands and the luminosity may be still close to the Eddington value, $\sim 10^{43} \ \rm erg \ s^{-1}$~\citep{Piran_et_al_2015}. 
Given the intergalactic absorption, such emissions from high-z events are difficult to be detected by on-going and up-coming transient surveys.

\subsection{Event rate}
We briefly estimate the rate of such high-z TDEs. 
The comoving number density of SMSs or DCBHs in the early growth is estimated to be $\sim 10^{-3} \rm \ Mpc^{-3} z^{-1}$ for $z \gtrsim 10$~\citep{Johnson_et_al_2013}, 
which corresponds to $\sim 100 \ \rm deg^{-2}$ of such systems in the sky. 
Given $\sim 10$ TDEs by each DCBH and a jet beaming factor of $f_{\rm b} \sim 0.01$, the total rate at $z \sim 10-20$ is estimated to be a few $\rm sky^{-1} yr^{-1}$.

Note that uncertainties of the DCBH number density are fairly large at this stage and can be significantly larger than the above value~\citep[see e.g.,][]{Agarwal_et_al_2012,Yue+14}.
Also, if the inward migration of the stars before joining the cluster is more efficient, the size of the star cluster becomes significantly smaller than the disk fragmentation radius,  
in which case the TDE rate is enhanced. 
On the other hand, relativistic jets may not accompany in some cases e.g., due to absence of large-scale magnetic field amplification in the TDE disk. 
In fact, off-axis radio afterglow observations of local TDEs suggest that only $\lesssim 10 \ \%$ of them are accompanied by powerful relativistic jets~\citep{van_Velzen_et_al_2013,Bower_et_al_2013}.

\vspace{1\baselineskip}
\section{Summary}
\label{sec:summary}
We analytically calculate properties of a nuclear accretion disk around a direct collapse black hole~(DCBH) within a few Myr after its formation.  
The outer disk is gravitationally unstable and fragments into clumps at $\sim 0.01-0.1 \ \rm pc$. 
The clumps evolve into Pop III stars with a typical mass of $\sim 10-100 \ M_\odot$, which will form a dens star cluster. 
The relaxation time of the cluster is estimated to be $\sim 10^{5}~\rm yr$ and shorter than the stellar lifetime of a few Myr.  
We can expect that $\sim 10$ of massive meal-poor stars are tidally disrupted by each DCBH. 
If a relativistic jet is launched by such a tidal disruption event, bright X-ray transients with a duration of 
a few months to $\sim$ yr could be produced and detectable by {\it Swift} BAT and eROSITA even from $z \sim 20$.  
Given a formation rate of DCBHs $\sim 10^{-3} \rm \ Mpc^{-3} z^{-1}$, the all sky event rate could be a few times per year, although the uncertainties are fairly large. 
Around the time when the DCBH mass is doubled, gas accretion will be strongly suppressed by various AGN feedback effects;  
the radiative and mechanical feedback from the inner slim disk, star cluster, and TDE jets, 
which can be probed by follow-up observations by eVLA and JWST.

\section*{Acknowledgements}
We thank Zolt$\rm \acute{a}$n Haiman, Nicholas Stone, Kunihito Ioka, and Amy Lien for valuable comments. 
KK is supported by NASA through Einstein Postdoctoral Fellowship grant number PF4-150123 awarded by the Chandra X-ray Center, 
which is operated by the Smithsonian Astrophysical Observatory for NASA under contract NAS8-03060.
KI is supported by the Simons Foundation through the Simons Society of Fellows.


\end{document}